\documentstyle[12pt]{article}
\begin{document}
\font\ninerm = cmr9

\def\footnoterule{\kern-3pt \hrule width \hsize \kern2.5pt}

\pagestyle{empty}

\begin{flushright}
CERN-TH/2000-035\\
hep-th/0001207 \\
$~$ \\
January 2000
\end{flushright}

\vskip 0.5 cm

\begin{center}
{\Large\bf On a possible quantum limit for the stabilization
of moduli in brane-world scenarios}
\end{center}
\vskip 1.5 cm
\begin{center}
{\bf Giovanni AMELINO-CAMELIA\footnote{{\it Marie Curie
Fellow} of the European Union
(address from February 2000: Dipartimento di Fisica,
Universit\'a di Roma ``La Sapienza'',
Piazzale Moro 2, Roma, Italy)}}
\end{center}
\begin{center}
{\it Theory Division, CERN, CH-1211, Geneva, Switzerland}
\end{center}

\vspace{1cm}
\begin{center}
{\bf ABSTRACT}
\end{center}

{\leftskip=0.6in \rightskip=0.6in

I consider the implications for brane-world scenarios 
of the rather robust quantum-gravity expectation that
there should be a quantum minimum limit
on the uncertainty of all physical length scales.
In order to illustrate the possible significance of this issue,
I observe that, according to a plausible estimate,
the quantum limit on the length scales that characterize the 
bulk geometry could affect severely the phenomenology of a
recently-proposed brane-world scenario.

}

\newpage
\baselineskip 12pt plus .5pt minus .5pt
\pagenumbering{arabic}
\pagestyle{plain} 


An extensive research effort (see, {\it e.g.}, 
Refs.~\cite{addlarge,bw,grwlarge,rs1,rs2,coka,lei,gfg} 
and references therein) has been recently devoted to
the possibility that the non-gravitational degrees of freedom 
be confined to one or more p-branes while 
gravitational degrees of freedom have access 
to some extra dimensions.
With respect to the development of related
formalism an important observation
is that in certain string theories it is quite natural~\cite{bw}
to obtain this type of different properties for gravitational and
non-gravitational degrees of freedom.
Concerning phenomenological implications, 
since it is the gravitational realm which is most affected
by these ``brane-world scenarios'', it is not surprising that
significant constraints come from the requirement that 
classical gravity should behave as observed in the regimes 
we have already explored experimentally.
On the quantum-gravity side some constraints also emerge;
in particular, interestingly,
while more conventional pictures lead
to graviton effects that are negligibly small, 
one finds~\cite{grwlarge}
that certain portions of the parameter
space of a given brane-world scenario turn out to be
excluded for predicting graviton effects that are inconsistent
with data obtained at existing particle colliders.
Larger portions of these parameter spaces will be probed
at planned colliders, such as LHC at CERN.

In this brief note I observe that, in addition to graviton
contributions to processes studied at particle colliders,
there is another class of quantum-gravity
effects which could have important implications for brane-world scenarios.
These effects are associated with 
the rather robust quantum-gravity 
expectation~\cite{wheely,venekonmen,padma,dharam94grf,gacmpla,dbrscatt}  
that physical length scales should not be definable with perfect accuracy,
there should be a minimum length uncertainty, 
and there should be quantum fluctuations of lengths.
This is conventionally (and somewhat generically) expressed with
formulas of the type $\Delta R \ge L_{min}$, intended to be
valid for any physical length scale.
I shall argue that, if 
such quantum limitations on the stabilization of length scales
apply to the length scales that characterize the bulk geometry,
there might be implications also for observables on
the brane where the Standard Model fields reside.

In conventional quantum-gravity 
scenarios~\cite{wheely,padma,dharam94grf,gacmpla}
$L_{min}$ is expected to coincide with $L_{QG}$,
the length scale that characterizes the strength
of gravitational interactions ($L_{QG}$ would be given by the
Planck length $L_p \sim 10^{-35}m$
in the conventional picture with only 3+1
space-time dimensions, but in the bulk of a brane-world scenario
one can have $L_{QG} \gg L_p$).

In quantum-gravity scenarios based on string theory
traditionally there has been the expectation~\cite{venekonmen}  
that the measurability bound should be even more
stringent: $L_{min} \sim L_s > L_{QG}$,
where $L_s$ is the string length ($L_s > L_{QG}$ in the 
perturbative regime).
More recently the analysis of certain stringy scenarios
with several length scales~\cite{dbrscatt}  
has suggested that in presence of appropriate hierarchies
of scales it may be possible to have
$L_{min} < L_{QG}$.
For example, in the scenario considered in Ref.~\cite{dbrscatt}
it appears that $L_{min} \sim (M_{D0} L_s)^{-1/12} L_{QG} < L_{QG}$,
where $M_{D0}$ is the mass of D-particles.

For brane-world scenarios in which quantum gravity 
(possibly in the guise of a string theory) 
behaves in the bulk in such a way that
$L_{min} \ge L_{QG}$
one would find that 
every given length scale $R_{bulk}$ characterizing the bulk geometry
({\it e.g.}, a curvature radius or an overall length of a finite
extra dimension)
would be affected by a quantum 
limitation: $\Delta R_{bulk} \ge L_{min} \ge L_{QG}$.
In the ordinary case, in which $L_{QG} \sim L_p$,
such quantum limits are very weak
for all lengths $R$ that we can access experimentally
(extremely small relative uncertainty $\Delta R/R \sim L_{QG}/R$),
but in the bulk of a brane-world scenario they can be significant
because $L_{QG} \gg L_p$ and some
of the length
scales $R_{bulk}$ are not much larger than $L_{QG}$.

In the mentioned stringy scenarios with several length scales
and an appropriate hierarchy of scales 
it might be possible to have
$\Delta R_{bulk} \sim L_{min} < L_{QG}$,
but values of $\Delta R_{bulk}$ that are significantly
smaller than $L_{QG}$ may require a strong hierarchy of scales.
For example, in the scenario considered in Ref.~\cite{dbrscatt}
even just the availability of $\Delta R_{bulk}$ of order, say,
$L_{QG}/1000$, would already require a very strong hierarchy
between the mass of D-particles and the string scale:
$M_{D0} \ge 10^{36}/L_s$.
The fact that the availability of $\Delta R_{bulk}$
significantly smaller than $L_{QG}$ may require such strong
hierarchies can be quite significant since most brane-world
scenarios intend to solve the ordinary ``hierarchy problem''
and may therefore loose most of their motivation if
requiring for other reasons (see below) 
that some new hierarchy problems arise.

Let me now discuss the possible implications of this set
of ideas in the significant illustrative example provided by
the model proposed by Randall and Sundrum in Ref.~\cite{rs1},
which in particular
assumes that all length
scales characterizing the bulk geometry are not too far from
the fundamental bulk-gravitational 
length scale $L_{QG}$ (which in the model \cite{rs1}
is taken to be close to the TeV scale).
In this model of Ref.~\cite{rs1} the mass $m$
of an ordinary Standard-Model
field and the mass scale $M_p = 1/L_p$ 
setting the strength (weakness) of gravity on the brane 
where the Standard Model fields reside
can be related through an exponential of 
the ratio between two of the length
scales characterizing the bulk
geometry: $m = M_p \cdot exp(- \pi R_{bulk,1}/R_{bulk,2})$
(in the notation of Ref.~\cite{rs1} $R_{bulk,1} = r_c$
and $R_{bulk,2} = 1/k$).
Values of $exp(\pi R_{bulk,1}/R_{bulk,2})$ close to $10^{15}$
are of interest for a solution of 
the ordinary hierarchy problem~\cite{rs1},
but if $R_{bulk,1}$ (and/or $R_{bulk,2}$)
is not much bigger than $L_{min}$ one would then 
predict a rather significant limitation\footnote{For example, 
the relation $m = M_p \cdot exp(- \pi R_{bulk,1}/R_{bulk,2})$
for $exp(\pi R_{bulk,1}/R_{bulk,2})=10^{15}$ 
and $R_{bulk,1} \sim 100 L_{min} \sim 100 \Delta R_{bulk,1}$
leads to $\Delta (m/M_p)$ of order $m/M_p$.}
on the accuracy of
the ratio $m/M_p$, while instead we measure 
with great accuracy both the masses of Standard Model particles
and the strength of ordinary gravitational interactions.

If the quantum gravity (or string theory) 
appropriate for the model of Ref.~\cite{rs1}
behaves in the bulk in such a way that
$L_{min} \ge L_{QG}$ the length scales 
$R_{bulk,1}$ and $R_{bulk,2}$
should indeed not be much bigger than $L_{min}$,
since, as mentioned,
in the model of Ref.~\cite{rs1} all length
scales characterizing the bulk geometry 
are not too far from $L_{QG}$.
In this case the alarming prediction
of significant limitations on the accuracy of
the ratio $m/M_p$ appears to be inevitable.

If the model of Ref.~\cite{rs1}
could be embedded in
a stringy quantum-gravity scenario 
of the type considered in Ref.~\cite{dbrscatt},
with several length scales
and a hierarchy of scales 
appropriate for having
$L_{min} \ll L_{QG}$,
one might be able to escape the prediction
of significant limitations on the accuracy of
the ratio $m/M_p$,
but then, as mentioned,
one would easily end up having a
new hierarchy problem associated with the
requirement $L_{min} \ll L_{QG}$.

In summary, at least at the heuristic level of the present
discussion, it would seem that the quantum-gravity expectation
that there should be a limit $\Delta R_{bulk} \ge L_{min}$
on the measurability of any length scale $R_{bulk}$
might affect non-trivially the analysis of
the scenario proposed in Ref.~\cite{rs1}.
Of course, a definite statement must await
more rigorous and quantitative analyses of the  
quantum properties of the bulk geometry
in models based on the scenario proposed in Ref.~\cite{rs1}.
The analyses should tell us whether 
$L_{min} < L_{QG}$ (actually, even though the evidence 
for an $L_{min} > 0$ is
quite robust~\cite{wheely,venekonmen,padma,dharam94grf,gacmpla,dbrscatt},
one cannot exclude that $L_{min} = 0$ might be found in some
quantum-gravity or stringy-quantum-gravity pictures)
or $L_{min} \ge L_{QG}$, and if $L_{min} < L_{QG}$ 
an estimate should be given of the amount of tuning
required to eliminate the ordinary hierarchy problem.

The example of the model proposed in Ref.~\cite{rs1} 
might also indicate that in general any claim of
consistency of a brane-world scenario must await 
the results of at least 
the level of analysis of the
quantum properties of the bulk geometry
necessary to address rigorously the issues I considered 
heuristically here.
It is important that such analyses be performed in
very physical terms, always resorting to
operative definitions of gravitational observables.
It is in fact well 
known~\cite{dharam94grf,gacmpla,bergstac} that formal 
estimates of quantum uncertainties in geometric observables
can be misleading.
For example, one finds~\cite{dharam94grf,gacmpla,bergstac}
that it is not sufficient to identify formally one of 
the objects in the formalism as a distance observable;
it is instead necessary to analyze an operative
definition of distance and consider all the
possible limitations which might be caused by each of the 
elements of the measurement procedure.
It is perhaps worth emphasizing that the operative
definition of gravitational observables, which is already
a delicate task in more conventional physical 
scenarios~\cite{gacmpla,bergstac,rovellimrs},
might be a formidable task in the case of those observables
of a brane-world scenario that concern the bulk geometry;
in fact, the measurement procedures that have been discussed 
in the conventional quantum-gravity
literature all rely~\cite{gacmpla,bergstac,rovellimrs}
on several non-gravitational elements, while the bulk
is not accessible to non-gravitational degrees of freedom.
It might be nontrivial even to establish what is 
genuinely observable~\cite{rovellimrs}
in the bulk, and what type of measurement procedures, particularly
with respect to the probes to be exchanged, 
would be appropriate.

Besides the analysis of 
possible quantum limits for the stabilization
of geometric observables in the bulk,
it might be also necessary~\cite{polonpap} to consider
quantum limits for the stabilization
of geometric observables 
of the brane where the Standard Model fields reside;
in fact, on some of these observables we start to have significant
experimental constraints~\cite{polonpap,gacgwi}.

\bigskip
\bigskip
\bigskip
\bigskip
It is a pleasure to thank Yaron Oz,
for conversations on the results reported in Ref.~\cite{dbrscatt},
and Gabriele Veneziano, for conversations on 
the results reported in Ref.~\cite{venekonmen}.

\baselineskip 12pt plus .5pt minus .5pt

\end{document}